\documentclass{ws-procs9x6_mod}

\newcommand{\be}{\begin{equation}}
\newcommand{\ee}{\end{equation}}
\newcommand{\ba}{\begin{eqnarray}}
\newcommand{\ea}{\end{eqnarray}}
\newcommand{\vev}[1]{\left\langle #1\right\rangle}
\newcommand{\lsp}{\tilde{\chi}}

\begin{document}

\title{RELAXATION DARK ENERGY \\ IN NON-CRITICAL STRING COSMOLOGIES AND \\ ASTROPHYSICAL DATA}

\author{NIKOLAOS~E.~MAVROMATOS~\footnote{Conference Speaker}}

\address{Physics Department, Theoretical Physics,
King's College London, \\
Strand, London WC2R 2LS, UK \\
E-mail: nikolaos.mavromatos@kcl.ac.uk}

\author{VASILIKI~A.~MITSOU}

\address{Instituto de F\'{i}sica Corpuscular (IFIC), CSIC --
Universitat de Val\`{e}ncia, \\
Edificio Institutos de Paterna, P.O.\ Box 22085, E-46071
Valencia, Spain \\
and \\
CERN, Physics Department, CH-1211 Geneva 23, Switzerland\\
E-mail: vasiliki.mitsou@cern.ch}

\begin{abstract}

In this talk we review briefly the basic features of non-critical (dissipative) String Cosmologies, and we confront some of these  models with supernova data.
We pay particular attention to the off-shell and dilaton
contributions to the dynamical evolution equations of the
non-critical string Universe, as well as the Boltzmann equation for species
abundances. The latter could have important consequences for the modification
of astrophysical constraints on physically appealing particle physics models,
such as supersymmetry. The data fits show that non-critical string cosmologies may be viable alternatives to $\Lambda$CDM model.

\end{abstract}


\bodymatter

\section{Introduction}\label{sec:intro}

There is a plethora of astrophysical evidence today, from supernovae
measurements~\cite{HST,SNLS}, the spectrum of fluctuations in the cosmic
microwave background~\cite{wmap}, baryon oscillations~\cite{baryon} and other
cosmological data, indicating that the expansion of the Universe is currently
accelerating. The energy budget of the Universe seems to be dominated (by more
than 70\%) at the present epoch by a mysterious dark energy component. Many theoretical models provide
possible explanations for the dark energy, ranging from a cosmological constant
\cite{concordance} to super-horizon perturbations \cite{riotto} and
time-varying quintessence scenarios \cite{steinhardt}, in which the dark energy
is due to a smoothly varying (scalar) field which dominates cosmology in the
present era. In the context of string theory \cite{gsw,polchinski}, the most
successful and mathematically complete theory of quantum gravity available to
date, such a time-dependent ``quintessence'' field is provided by the scalar
dilaton field of the gravitational string multiple~\cite{aben,gasperini,emnw}.
The current astrophysical data are capable of placing severe constraints on the nature of the dark energy, whose equation of state may be determined by means
of an appropriate global fit. Most of the analyses so far are based on
effective four-dimensional Robertson-Walker Universes, which satisfy on-shell
dynamical equations of motion of the Einstein-Friedman form. Even in modern
approaches to brane cosmology, which are described by equations that deviate
during early eras of the Universe from the standard Friedman equation (which is linear in the energy density), the underlying dynamics is assumed to be of
classical equilibrium (on-shell) nature, in the sense that it satisfies a set
of equations of motion derived from the appropriate minimization of an
effective space-time Lagrangian.

However, cosmology may not be an entirely classical equilibrium situation.  The
initial Big Bang or other catastrophic cosmic event, such as a collision of two
brane worlds in the modern approach to strings~\cite{polchinski}, which led to
the initial rapid expansion of the Universe, may have caused a significant
departure from classical equilibrium dynamics in the early Universe, whose
signatures may still be present at later epochs including the present era. One
specific model for the cosmological dark energy which is of this type, being
associated with a rolling dilaton field that is a remnant of this
non-equilibrium phase, was formulated~\cite{emnw,diamandis,diamandis2} in the
framework of non-critical string theory~\cite{aben,ddk,emn}. We call this
scenario `Q-cosmology'. It is of outmost importance to confront the currently
available precision astrophysical data with such non equilibrium stringy
cosmologies. The central purpose of this talk is to present a first step
towards this direction, namely a confrontation of cosmological data on
high-redshift supernovae~\cite{emmn} with Q-cosmologies and compare the results
with the predictions of the conventional $\Lambda$CDM model \cite{concordance}
and the super-horizon model~\cite{riotto}. Care must be taken in interpreting
the Q-cosmology scenario. Since such a non-equilibrium, non-classical theory
\emph{is not described by the equations of motion derived by extremising an
effective space-time Lagrangian}, one must use a more general formalism to make
predictions that can be confronted with the current data.

\section{Dissipative Q-Cosmology Basics}

\subsection{Off-shell (non-critical-string) terms and dissipative
cosmological equations}

Q-cosmology are stringy cosmological models, based on the (perturbative)
formulation of $\sigma$-models propagating mostly on dilaton and graviton
backgrounds, which, however, are not world-sheet conformal invariant. The
deviation from conformal invariance may be due to a number of reasons, for
instance cosmically catastrophic events at an early stage of the Universe
history, such as Big Bang, the collision of two (or more) brane worlds {\it
etc.} An important formal ingredient of this approach is the
identification~\cite{emn} of target time with the zero mode $\rho$ of the
Liouville field~\cite{ddk}, which is an extra world-sheet field, introduced in
order to restore the conformal invariance of the world-sheet theory. The
dynamics of this latter identification is encoded in the solution of the
generalized conformal invariance conditions, after Liouville dressing, which
read in the $\sigma$-model frame~\cite{emn,ddk}:
\begin{equation}
-{\tilde \beta^i } = {g^i}'' + Q{g^i}', \label{liouveq}
\end{equation}
where the prime denotes differentiation with respect to $\rho$, and the overall
minus sign on the left-hand side pertains to supercritical strings~\cite{aben},
with a timelike signature of the Liouville mode, for which the central charge
deficit $Q^2 > 0$ by convention. In physical terms, $Q$ depends on the
microscopic details of the specific non-equilibrium string theory model at
hand. For instance, in the case of adiabatic brane world collisions with
bouncing (recoiling) worlds~\cite{gravanis,brany,emnw}, $Q^2 \propto v^4 $,
where $v \ll 1$ is the (relative) recoil velocity of the brane worlds after the
collision. Notice the dissipation, proportional to the (square root) of the
central charge deficit $Q$, on the right-hand side of \eref{liouveq}, which
heralds the adjective \emph{Dissipative} to the associate
non-critical-string-inspired Cosmological model. Moreover, the Weyl anomaly
coefficients ${\tilde \beta^i },~i= \{\Phi, g_{\mu\nu} \}$, whose vanishing
would guarantee local conformal invariance of the string-cosmology background,
are associated with off-shell variations of a low-energy effective
string-inspired action, ${\cal S}[g^j]$:
\begin{equation}
\frac{\delta {\cal S}[g] }{\delta g^i} = {\cal G}_{ij}{\tilde \beta}^j, \qquad
{\cal G}_{ij}= {\rm Lim}_{z \to 0} z^2 {\overline z}^2 \langle V_i(z,{\overline
z}) V_j(0,0)\rangle,
 \label{offshell}
\end{equation}
with $z,{\overline z}$ (complex) world-sheet coordinates, ${\cal G}_{ij} $ the
Zamolodchikov metric in theory space of strings~\cite{gsw,emn}, and $V_i$ the
$\sigma$-model vertex operators associated with the $\sigma$-model background
field $g^i$. It is this off-shell relation that characterises the entire
non-critical ($Q$) Cosmology framework, associated physically with a
non-equilibrium situation as a result of an initial cosmically catastrophic
event, at the beginning of the (irreversible) Liouville/cosmic-time flow.

The detailed dynamics of \eref{liouveq} are encoded in the solution for the
scale factor $a(t)$ and the dilaton $\Phi(t)$ in the simplified model
considered in Ref.~\refcite{diamandis2}, after the identification of the
Liouville mode with the target time. In fact, upon the inclusion of matter
backgrounds, the associated equations, after compactification to four
target-space dimensions, read in the so-called Einstein frame (i.e., an
appropriate redefinition of the $\sigma$-model graviton, such that the
graviton-dilaton effective action has a
canonically normalised Einstein curvature scalar term~\cite{aben}): %
\ba &&3  H^2 - {\tilde{\varrho}}_m - \varrho_{\Phi}\;=\; \frac{e^{2 \Phi}}{2}
\tilde{\cal{G}}_{\Phi}, \quad 2\dot{H}+{\tilde{\varrho}}_m + \varrho_{\Phi}+
{\tilde{p}}_m +p_{\Phi}\;=\; \frac{\tilde{\cal{G}}_{ii}}{a^2}, \nonumber  \\
&& \ddot{\Phi}+3 H \dot{\Phi}+ \frac{1}{4} \; \frac{\partial {
\hat{V}}_{all} }{\partial \Phi} + \frac{1}{2} \;(
{\tilde{\varrho}}_m - 3 {\tilde{p}}_m )= - \frac{3}{2}\; \frac{
\;\tilde{\cal{G}}_{ii}}{ \;a^2}- \, \frac{e^{2 \Phi}}{2} \;
\tilde{\cal{G}}_{\Phi}, \nonumber \\
&& \tilde{\cal{G}}_{\Phi} \;=\; e^{\;-2 \Phi}\;( \ddot{\Phi} - {\dot{\Phi}}^2 +
Q e^{\Phi} \dot{\Phi}), \nonumber \\ && \tilde{\cal{G}}_{ii} = 2 a^2 (
\ddot{\Phi} + 3 H \dot{\Phi} + {\dot{\Phi}}^2 + ( 1 - q ) H^2 + Q e^{\Phi} (
\dot{\Phi}+ H ))~, \label{eqall} \ea %
where ${\tilde{\varrho}}_m$ (${\tilde{p}}_m $) denotes the matter energy
density (pressure), including dark matter contributions, and $\varrho_{\Phi}$
($p_\Phi$) are the corresponding quantities for the dilaton dark-energy fluid.
The overdots in the above equations denote derivatives with respect to the
Einstein time, which is related to the Robertson-Walker cosmic time $t_{RW}$ by
$t=\omega \;t_{RW}$. Without loss of generality we have taken $\omega =
\sqrt{3} H_0$ where $H_0$ is the present-day Hubble parameter. With this choice
for $\omega$ the densities appearing in Eqs.~(\ref{eqall}) are given in units
of the critical density. In the above equations, $H=\dot a/a$ is the Hubble
parameter and $q$ is the deceleration parameter of the Universe $q \equiv -
\ddot{a} a / {\dot{a}}^2$, and are both functions of (Einstein frame) cosmic
time. The potential ${\hat{V}}_{all}$ appearing above is defined by
${\hat{V}}_{all}= 2 Q^2 \exp{(2 \Phi)} + V$ where, in order to cover more
general cases, we have also allowed for a (e.g., string-loop-induced) potential
term in the four-dimensional action,  $- \int d^4y \sqrt{-G}V$. We have also
assumed a (spatially) flat Universe. The dilaton energy density and pressure
are given in this class of models by:~ $\varrho_{\Phi} = \frac{1}{2} (
2{\dot{\Phi}}^2+{\hat{V}}_{all} ),~ p_{\Phi} = \frac{1}{2} ( 2
{\dot{\Phi}}^2-{\hat{V}}_{all} ).$ The dependence of the central charge deficit
$Q(t)$ on the cosmic time stems from the running of the latter with the
world-sheet RG scale~\cite{emn,brany}, and is provided by the Curci - Paffuti
equation~\cite{curci} expressing the renormalisability of the world-sheet
theory. To leading order in an $\alpha '$ expansion, which we restrict
ourselves here, this equation in the Einstein frame
reads~\cite{diamandis2}:~$\frac{d \tilde{\cal{G}}_{\Phi} }{d t_E} = - 6 e^{ -2
\Phi} ( H + \dot{\Phi} )  \frac{\tilde{\cal{G}}_{ii}}{a^2}.$ For completeness
we also state here the continuity equation of the matter stress tensor, which
is not an independent equation, but can be obtained from \eref{eqall} by
appropriate algebraic manipulations: $\frac{d{\tilde{\varrho}}_m }{dt_E} + 3 H
( {\tilde{\varrho}}_m +{\tilde p}_m) + \frac{\dot Q}{2} \frac{\partial {
\hat{V}}_{all} }{\partial Q} - \dot{\Phi} ({\tilde{\varrho}}_m - 3
{\tilde{p}}_m ) = 6 (H+\dot{\Phi}) \frac{ \tilde{\cal{G}}_{ii}}{a^2}.$

A consistent numerical solution of $a(t)$, $\Phi(t)$, and the various
densities, including back reaction of matter onto the space-time geometry, has
been discussed in Ref.~\refcite{diamandis2}, where we refer the interested
reader for further study. We note the existence of exotic matter scaling in
this approach today, which no longer scales as dust, but it includes different
scaling components, e.g. $a^{-\delta}$, with $\delta $ close to 4. We 
quote below the final result of our parametrisation for $H(z)$ in the
Q-cosmology framework, at late eras,
such as the ones pertinent to the
supernova and other data ($0<z<2$),
where some analytic approximations are
allowed~\cite{emmn}:
\begin{equation}\label{formulaforfit}
\frac{H(z)}{H_0} = \sqrt{{\Omega }_3 (1 + z)^3 + {\Omega }_{\delta} (1 +
z)^\delta + {\Omega}_2(1 + z)^2}~,~~{\Omega }_3 + {\Omega}_{\delta} + {\Omega}_2 = 1~,
\end{equation}
with the densities $\Omega_{2,3,\delta}$ corresponding to present-day values
($z = 0$). We stress, however, that a complete analysis of the non-critical and dilaton effects, which turn out to be important in the present era after
the inclusion of matter, requires numerical treatment~\cite{diamandis2}.
In view of this,
the exponent
$\delta$ is treated from now on as a fitting parameter, and in fact it {\it may
even be} $z$-dependent to cover more general cases, and get agreement with the
numerical treatment of Ref.~\refcite{diamandis2}.
It is important to note that the various $\Omega_i$ contain contributions from
\emph{both} dark energy and matter energy densities. As explained in detail in
Ref.~\refcite{emmn}, $\Omega_3$ does not merely represent ordinary matter
effects, but also receives contributions from the dilaton dark energy. In fact,
the sign of $\Omega_3$ depends on details of the underlying theory, and it
could even be \emph{negative}. For instance, Kaluza-Klein graviton modes in
certain brane-inspired models \cite{kaluza} yield negative dust contributions.
In a similar vein, the exotic contributions scaling as $(1+z)^\delta$ are
affected by the off-shell Liouville terms of Q-cosmology. It is because of the
similar scaling behaviours of dark matter and dilaton dark energy that we
reverted to the notation $\Omega_i$, $i=2,3,\delta,$ in \eref{formulaforfit}.
To disentangle the ordinary matter and dilaton contributions one may have to
resort to further studies on the equation of state of the various components,
which we do not study here or in \cite{emmn}. More generally, one could have included
a cosmological constant $\Omega_\Lambda$ contribution in \eref{formulaforfit},
which may be induced in certain brane-world inspired models. We do not do so in
this work, as our primary interest is to fit Q-cosmology models
\cite{emnw,diamandis,diamandis2}, which are characterised by dark energy
densities that relax to zero.

\subsection{Dilaton and off-shell modifications to cosmic evolution of species
abundances}

The above formalism can be used to derive the dilaton and non-critical-string
effects on Boltzmann-type evolution equations~\cite{kolb} of a number of
densities for certain particle physics species, playing the r\^ole of dominant
dark matter candidates. In the presence of (time-dependent) dilaton and
off-shell graviton fields, that couple to matter, the phase-space density of
species depends on sources, $f(|\vec p\,|, t, \Phi (t,\rho),
g_{\mu\nu}(t,\rho))$, where $p$ is the momentum, and  $\rho$ is the Liouville
mode, which eventually is going to be identified with a function of the target
time $t$~\cite{emn}, whose form is dictated by both conformal-field-theory and
target-space dynamical considerations, such as the minimisation of the
effective potential~\cite{gravanis}. A detailed analysis~\cite{lmnbolt} shows
that the Boltzmann equation for the density of species $X$, with mass $m_X$,
assumed to be a dominant dark matter candidate, $n = \int d^3\!p\,f$, can be
written in a compact form that represents collectively the
dilaton-dissipative-source and
non-critical-string contributions as external-source $\Gamma (t) n$ terms: %
\ba %
\frac{d n}{dt} + 3\frac{\dot a}{a}n = \Gamma (t)n - \vev{v \sigma} ( n^2 -
n_{eq}^2)~, \quad \Gamma (t) \equiv  \dot \Phi  + \frac{1}{2}\eta
e^{-\Phi}g^{\mu\nu}{\tilde \beta}_{\mu\nu}^{\rm Grav}, \label{boltzfinal}
\ea %
where the last term on the right-hand-side of the Boltzman equation
is the collision term,
expressed in terms of the thermal average of the
cross section $\sigma$ times the M\"{o}ller velocity $v$ of the annihilated
particles~\cite{kolb};
$\eta$ is a world-sheet
renormalization-scheme dependent parameter~\cite{emn,lmnbolt}; for our purposes
here we work in the physical scheme $\eta = -1$.
To find an explicit expression for $\Gamma (t)$ in our case requires
the full solution of the Q-cosmological equations~\cite{diamandis2}.
Depending on the sign of
$\Gamma (t)$ one has different effects on the relic abundance of the species
$X$ with density $n$, which we now proceed to analyse briefly.
Before the decoupling (`freeze out') time $t_{f}$, $t < t_f$, equilibrium is
maintained and thus $n=n_{eq}$ for such an era. However, it is crucial to
observe that, as a result of the presence of the source $\Gamma$ terms,
$n_{eq}$ no longer scales with the inverse of the cubic power  of the expansion
radius $a$, which was the case in conventional (on-shell) cosmological models.
Under some plausible assumptions, for instance that the entropy remains
approximately constant after the freeze out temperature, and that one can
define a temperature $T$ for the ordinary matter degrees of freedom, but {\it
not} for the ones pertaining to (dilatonic) dark energy, we can solve the
Boltzmann equation and arrive at the dilaton and non-critical string
corrections to the freeze out temperature of, say, the lightest supersymmetric
particle, $\lsp$, assumed to be the dominant dark matter candidate (we define
$x \equiv T/m_{\lsp}$):%
\ba x_f^{-1} = \ln \left[ \left(0.03824 g_s \frac{M_{\it Planck}~
m_{\lsp}}{\sqrt{g_{*}}} x_f^{1/2} {\vev{ v \sigma}}_f \right)\!\left(
\frac{g_{*}}{ {\tilde{g}}_{*} }\right)^{1/2} \right] + \int_{x_f}^{x_{in}}
\frac{\Gamma dx}{H x}, \label{frpo}
\ea %
whereby ${\tilde{g}}_{\it eff}$ is simply defined by~\cite{lmnbolt}
$\varrho+\Delta \varrho \equiv \frac{\pi^2}{30} T^4 {\tilde{g}}_{\it eff} $.
$\Delta \varrho$ incorporates the effects of the additional contributions due
to the non-critical (off-shell) terms and the dilaton dissipative source, which
are not accounted for by the ordinary matter degrees of freedom $g_{\it eff}$
of conventional Cosmology~\cite{kolb}. Recalling that only the degrees of
freedom involved in ordinary matter energy density $\rho$ are
thermal~\cite{kolb}, $\rho= \frac{\pi^2}{30} T^4 g_{\it eff}(T) $, whilst from
the dynamical equations of Q-cosmology $H^2=\frac{8 \pi G_N}{3} (\rho + \Delta
\rho) $, we obtain~\cite{lmnbolt} ${\tilde g}_{\it eff}=g_{\it eff} +
\frac{30}{\pi^2} \, T^{-4} \Delta \rho$.  We can also assume that the
${\lsp}$'s decoupled before neutrinos. Then, for
the relic abundance, we derive the following approximate result~\cite{lmnbolt}%
\ba \Omega_{\tilde{\chi}} h_0^2 = \left( \Omega_{\tilde{\chi}} h_0^2
\right)_{\it no-source} \times {\left(  \frac{{\tilde g}_{*}}{g_{*}}
\right)}^{1/2} \; {\exp}\left( \int_{x_{0}}^{x_f}  \frac{\Gamma dx}{H x}
\right) ,
\label{relic} \ea %
with $\left( \Omega_{\tilde{\chi}} h_0^2 \right)_{\it no-source} = \frac{1.066
\times  10^9  {\rm GeV}^{-1}}{ M_{\it Planck}  \sqrt{g_{*}}  J} $, ~$J\equiv
\int_{x_0}^{x_f} \vev{ v \sigma} dx$, and $x_f$ determined by \eref{frpo}. The
merit of casting the relic density in such a form is that it clearly exhibits
the effect of the presence of the source. It is immediately seen from
\eref{relic} that, depending on the sign of the source $\Gamma$, one may have
increase or reduction of the relic density as compared with the corresponding
value in the absence of the source. In this way, the constraints imposed on
supersymmetric extensions of the Standard Model from astrophysical
data~\cite{susyconstr}, such as those from WMAP-satellite measurements,
\cite{wmap} {\it etc.}, based on the available bounds of, say, cold dark matter
relic densities, need to be revisited. This is left for the future.

\section{Confronting Q-Cosmologies with Supernova Data}

We use the type-Ia supernovae (SN) data reported in Refs.~\refcite{HST,SNLS},
which are given in terms of the distance modulus $\mu = 5\log d_L + 25$, where
the luminosity distance $d_L$ (in megaparsecs) is related to the redshift $z$
via the Hubble rate $H$:
\begin{equation}\label{luminositydistanceredshift}
d_L = c(1 + z) \int_0^z \frac{dz'}{H(z')}.
\end{equation}
We note that this observable depends on the expansion history of the Universe
from $z$ to the present epoch, and recall that most of the available supernovae
have $z < 1$, although there is a handful with larger values of $z$. In the
analysis that follows, the predictions of the following three cosmological
models are investigated:
\begin{description}

  \item[\boldmath $\Lambda$CDM model\unboldmath] assuming a flat universe, where
  the Hubble rate is
  \begin{equation}
H(z) = H_0 \left(\Omega_{\rm M} ( 1 + z)^3 + \Omega_\Lambda ( 1 + z )^{3(1 +
w_0)}\right)^{1/2}.\label{hubble}
\end{equation}

  \item[Super-horizon model] in which the Hubble parameter is given by \cite{riotto}
  \begin{equation}
H(z) = {\overline a}^{-1}\frac{d{\overline a}(t)}{dt}= \frac{{\overline H}_0}{1
- \Psi_{\ell 0}}\left(a^{-3/2} - a^{-1/2}\Psi_{\ell 0}\right),
\label{modhubble}
\end{equation}
where $(1 + z)^{-1} = \overline{a}(t)$ and $\Psi_{\ell 0}$ is a free parameter.

  \item[Q-cosmology] discussed in the previous sections, with $H$ given by
  \eref{formulaforfit} and three parameters to be determined by the fit:
  $\Omega_3$, $\Omega_\delta$ and $\delta$.

\end{description}
Measurements are available of the distance moduli of 157~supernovae in a
so-called `gold' sample \cite{HST}, observed by ground-based facilities and the
Hubble Space Telescope (HST). In addition, a so-called `silver' dataset of
29~SN~1a is also available, of a slightly lesser spectrometric and photometric
quality. The analysis was also repeated including the `silver' supernovae data
---with a total of 186~SN--- with results comparable to those with the
`gold' dataset, proving the robustness of the analysis. Furthermore, we
analysed the measurements of 71~other high-redshift supernovae by the Supernova
Legacy Survey (SNLS) \cite{SNLS}, which are accompanied by a reference sample
of 44~nearby SN~Ia, yielding a total of 115~data points. Since the fits to the
two datasets are quite compatible \cite{emmn}, a combined analysis was finally
performed.

\begin{figure}
\begin{center}
\psfig{file=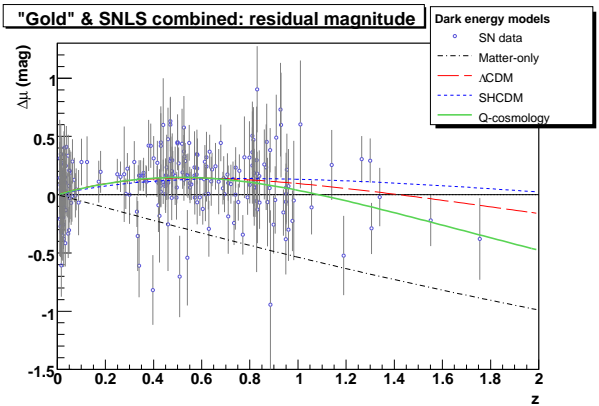,width=0.7\linewidth} 
\end{center}
\caption{Residual magnitude versus redshift for supernovae from the `gold' and
the SNLS datasets. Predictions of cosmological models for the best-fit
parameter values are superimposed: (i) Empty Universe; (ii) Universe with
matter only, $\Omega_{\rm M}=1$; (iii) $\Lambda$CDM model; (iv) super-horizon
model; and (v) off-shell Q-cosmology model. } \label{fig:SNres_comb}
\end{figure}

For illustration purposes, both data and predictions of cosmological models are
expressed in the following as residuals, $\Delta\mu$, from the empty-Universe
($\Omega_{\rm M}=0$) prediction. The combined sample of the `gold' (157~SN)
plus the SNLS sample (71~SN), yielding a sample of 228~supernovae in total, is
shown in \fref{fig:SNres_comb}, where the predictions of the cosmological
models under study are also displayed for the best-fit parameter values.

The analysis involves minimisation of the standard $\chi^2$ function with
respect to the cosmological model parameters. The best-fit parameter values
acquired by combining the `gold' sample and the SNLS supernovae datasets, the
$1\sigma$ errors and the corresponding $\chi^2$ values are listed in
\tref{tb:chi2} for the three cosmological models.
\begin{table}
\tbl{Parameter values favoured by the `gold' + SNLS combined data for
various cosmological models.}%
{\begin{tabular}{@{}lccc@{}}\toprule
Models & Best-fit parameters & $\chi^2$ & $\chi^2/{\rm dof}$ \\
\colrule
$\Lambda$CDM    & $\Omega_{\rm M}=0.274\pm0.017$    & 239   & 1.05 \\
Super-horizon   & $\Psi_{\ell0}=-0.87\pm0.06$       & 245   & 1.09 \\
                & $\Omega_3=-3.7\pm1.1$             &       &   \\
Q-cosmology     & $\Omega_{\delta}=1.3\pm0.7$       & 237   & 1.05 \\
                & $\delta=3.9\pm0.3$                &       &   \\
\botrule
\end{tabular}
}
\label{tb:chi2}
\end{table}

It is evident from \fref{fig:SNres_comb} and \tref{tb:chi2} that the standard
$\Lambda$CDM model fits the supernova data \emph{very well}, as expected from
earlier analyses. The super-horizon dark matter model also fits the supernova
data \emph{quite well}. Both of these models are {\it on-shell}, i.e., they
satisfy the pertinent Einstein's equations. Moreover, {\it off-shell} cosmology
models are also compatible with the data. As we discussed above, off-shell
effects are important in our Q-cosmology model. Introducing the appropriate
parametrisation (\ref{formulaforfit}) to allow for these off-shell effects, we
find that \emph{the Q-cosmology model may fit the supernovae data as well as
the standard $\varLambda$CDM model}.

\begin{figure}
\begin{center}
\psfig{file=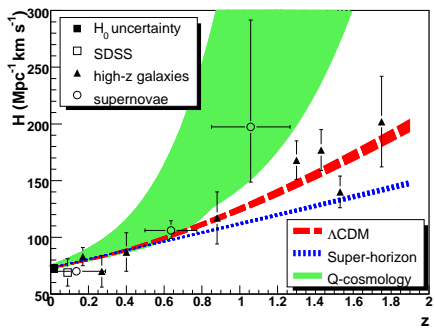,width=0.55\linewidth,clip=}
\end{center}
\caption{The Hubble parameter redshift relation for various cosmological models
and observational data. The bands represent 68\% confidence intervals derived
by the SN analysis of Ref.~\refcite{emmn} for the standard $\Lambda$CDM, the
super-horizon and the Q-cosmology models. The black rectangle shows the WMAP3
\cite{wmap} estimate for $H_0$, the squares show the measurements from SDSS
galaxies \cite{H_SDSS}, the triangles result from high-$z$ red galaxies
\cite{H_high_z}, and the circles correspond to a combined analysis \cite{HST06}
of supernovae data from Refs.~\refcite{HST} and~\refcite{SNLS}.}
\label{fig:HubbleCL}
\end{figure}

In \fref{fig:HubbleCL}, the Hubble parameter predictions of the various models
are compared with the data available from high-$z$ red galaxies
\cite{H_SDSS,H_high_z} and from sets of type~Ia SN \cite{HST06}. The bands
correspond to the 68\% confidence intervals deduced by the aforementioned
analysis of 228~supernovae (cf.\ \tref{tb:chi2}). The SN data constrain $H$
only up to $z\simeq1.3$ with a 50\% uncertainty, in contrast to the red
galaxies data, which provide stringent measurement of $H$ until $z\simeq1.8$.
The $\Lambda$CDM model is compatible with the data, however the super-horizon
model deviates from both galaxies and supernovae data. Q-cosmology, on the
other hand, predict higher values of $H$ than the galaxies data measurements,
though fully compatible with the SN data, as expected. Nevertheless, this
discrepancy is hoped to be remedied by repeating the analysis of
Ref.~\refcite{emmn} with the inclusion of a sample of 21~type-Ia supernovae
---including 13 with $z>1$--- recently released by HST \cite{HST06}. Moreover,
as remarked earlier, a $z$-dependent-$\delta$ parametrisation of $H(z)$
(\ref{formulaforfit}) may be necessary, as becomes clear from the highly
non-linear form of the numerical solution of some Q-cosmology models presented
in Ref.~\refcite{diamandis2}. In fact, fitting the numerical solution itself
with the data is probably the most complete treatment. We hope to tackle such
issues in the future.

For standard on-shell cosmologies, complementary information is provided by the
data on baryon acoustic oscillations \cite{baryon}, which show up in the
galaxy-galaxy correlation function at $z \sim 0.35$. However, for reasons
discussed in Ref.~\refcite{emmn}, the application of such an analysis to the
off-shell Q-cosmology model is an open issue, since the underlying theoretical
framework needs to be re-evaluated.

\section{Conclusions}

In this talk, we briefly reviewed the main predictions of Q-cosmologies,
including a brief discussion about the time-dependent-dilaton and
non-critical-string effects on the Boltzmann equation for species
abundances~\cite{lmnbolt}, which could lead to important modifications of the
constraints imposed by astro-physical data on interesting particle physics
models, such as supersymmetric extensions of the standard model. Then, we
discussed some initial steps towards demonstrating~\cite{emmn} that the
available supernova data are compatible with such non-critical-string-based
cosmologies \cite{emnw,diamandis2}, thus implying the possibility that such
models may be viable alternatives to the Standard $\Lambda$CDM model. As more
precision astrophysical data are coming into play, more stringent constraints
can be imposed on our non-critical string Q-cosmologies.

\vspace*{0.3cm} \noindent {\bf Acknowledgements: } We thank G.~Diamandis,
J.~Ellis, B.~Georgalas, A.~Lahanas and D.~Nanopoulos for the enjoyable
collaboration and discussions. The work of N.E.M.\ was supported in part by the
European Union through the Marie Curie Research and Training Network
\emph{UniverseNet} (MRTN-CT-2006-035863).

\bibliographystyle{ws-procs9x6}
\bibliography{ws-pro-sample}

\end{document}